\documentclass[aps,prl,preprint,superscriptaddress]{revtex4}

\begin{document}


\title{Dark energy and Chern-Simons like gravity from a dynamical four-form}


\author{Patrick Das Gupta}
\email[]{patrick@srb.org.in}
\affiliation{Department of Physics and Astrophysics, University of Delhi, Delhi - 110 007 (India)}


\date{\today}

\begin{abstract}
 We consider the dynamics of a four-form field $\tilde {w} $,  
 treating it as a distinct physical degree of freedom, independent of the 
 metric.  The equations of motion are derived from an action which, besides having the standard Hilbert-Einstein term and the matter part, consists of a new action for $\tilde {w} $. 
  The evolution of this four-form in the framework of a flat FRW model is studied, and it is shown that the parameters of the theory  admit solutions wherein it is possible to have
  an equation of state $p_\phi \approx -\epsilon_\phi $ for $\tilde {w} $, so that  it leads to an accelerating universe.  Taking cue from the paper by Jackiw and Pi (2003), we also
  put forward electromagnetic as well as gravitational `Chern-Simons' like terms using $\tilde {w} $ 
   that arise naturally in 4D  without any loss of Lorentz invariance. This entails on one hand a modified Einstein-Maxwell equation, having the potential to be contrained observationally by
  CMBR  and other astrophysical data, and an enlarged system of Einstein equation, on the other, involving a Cotton tensor.  It is shown that the presence of gravitational Chern-Simons like 
  term in the theory does not affect the flat FRW model analysis of the evolution of $\tilde {w} $.
  We also demonstrate that the scalar-density associated with $\tilde {w} $ can be
  employed to  construct a  generalized exterior derivative that converts a 
  p-form density to  a (p+1)-form density of  identical 
 weight.  
\end{abstract}


\maketitle

\section{}
There is growing evidence over almost a decade and a half that the rate of expansion of the universe is increasing with time \cite {Perl, Schmi, Riess98}. 
 A recent analysis with 414 Type Ia supernovae suggests that the equation of state parameter w for the hypothesized dark energy (DE), responsible for the
 acceleration of the expansion, is -0.969 with an error of about  10$\%$ \cite {Kowal}. The observed value of w is tantalizingly close to -1 suggesting a non-zero cosmological constant as the panacea.
 The cosmological constant scenario, however,  is replete with problems of fine-tuning and cosmic-coincidences \cite {Wein89, Vilen}. Alternate popular models of DE involve scalars like quintessence and 
 phantom fields \cite {Ratra88, Wett, Cald, Zlat, Dutta08, Cald1}. In this paper, we explore the possibility of generating an accelerating universe by means of a dynamical four-form field $\tilde {w}$ evolving in the 
 standard  4D spacetime.
 
 Four-forms derived from three-form gauge potentials have been invoked to examine the origin of cosmological constant from a different
 perspective \cite{Auri78, Duff, Auri80, Hawk, Brown, Duff1, Duncan, Turok, Auri04, Wu, Klink}. Related approaches have also been
 extended, in recent times, to studies pertaining to DE and the cosmological constant in the realm of string and M-theories where coupling of four-form fields with 
 branes is generic \cite{Bousso, Feng, Dvali, Vilen, Garr, Kall, Wohl, Garr1, Jarv}.  A dynamical volume-form independent of the metric has also been discussed 
 in the literature wherein the new degree of freedom is associated with a massless dilatonic scalar field that leads to a scalar-tensor theory of gravitation \cite {Graf, Graf1}.
Guendelman and Kaganovich (2008) have proposed a two-measures field theory in which besides the
 standard volume-measure $\sqrt {-g} d^4x$ there is a dynamical volume-form made up of four one-forms, each associated with an independent scalar field \cite {Guen}. Possibility 
 of  generating an inflationary phase
 by means of n-forms has also been studied recently \cite {Mota}.
 Contrary to the
 previous investigations, in the present 
 analysis we adopt a different approach treating  the scalar-density of weight +1 corresponding to the four-form $\tilde {w} $
 as fundamental, whose dynamics is determined by a Lagrangian density constructed out of its covariant derivatives.

In special relativity, there are two nontrivial tensors that are invariant under proper Lorentz 
transformations - the Minkowski metric $\eta_{\mu \nu}$ and the totally antisymmetric Levi-Civita tensor
$\epsilon_{\mu \nu \rho \sigma}$ \cite{Wein72}. When one makes a transition to classical 
general relativity, the metric encoding the geometry of the spacetime, attains a dynamical status $g_{\mu \nu} (x)$ whose time
 evolution is determined by the Einstein equation. Taking the analogy of $\eta_{\mu \nu}$ metamorphosising into a 
 dynamical $g_{\mu \nu} $ one step lateral, we put to scrutiny the hypothesis of Levi-Civita tensor transforming into a
  dynamical field  $w_{\mu \nu \rho \sigma} (x)$.

We propose that a four-form field $\tilde {w} $ which, in a coordinate basis, can be expressed as,
$$\tilde {w} = {1\over {4!}} w_{\mu \nu \rho \sigma}  \tilde{d} x^\mu \wedge   \tilde{d} x^\nu \wedge   \tilde{d} x^\rho \wedge  \tilde{d} x^\sigma $$
 represents a new physical degree of freedom in the gravitational theory, completely independent of 
 the metric, that couples universally to all fields.
Since  $w_{\mu \nu \rho \sigma} (x)$ is totally antisymmetric, it can be equated in a 4D spacetime manifold to $\phi (x) \epsilon_{\mu \nu \rho \sigma}$, where
$\phi (x) $ is a scalar-density of weight +1, with  $\phi \rightarrow \phi /J $,  under a general coordinate transformation with Jacobian $J$.
A  p-vector {\bf {w}} with totally antisymmetric components $w^{\mu \nu \rho \sigma}$ corresponding to the four-form $\tilde {w}$ can be constructed 
by demanding that $w^{\mu \nu \rho \sigma} \ w_{\mu \nu \rho \sigma}= - 4!$ \cite{Schutz80}. Then, it follows that $w^{\mu \nu \rho \sigma}= \epsilon^{\mu \nu \rho \sigma}/\phi(x)$. 
We assume $\phi = w_{ 0 1 2 3}$ to be dimensionless. 

The metric has not entered the picture thus far, but it does so as soon as  one switches on the gravitational interaction. The covariant derivatives of the four-form field and its corresponding
p-vector are given by,
$$w_{\mu \nu \alpha \beta \ ; \lambda} = [(\ln \phi)_{, \lambda} - \Gamma ^\sigma _{\sigma \lambda}] w_{\mu \nu \alpha \beta}\eqno(1)$$
and
$$w^{\mu \nu \alpha \beta}_{\ \ ; \lambda} = - [(\ln \phi)_{, \lambda} - \Gamma ^\sigma _{\sigma \lambda}] w^{\mu \nu \alpha \beta},\eqno(2)$$
which, by virtue of equivalence principle, guide us to write down below, an action $S$ that is invariant under 
general coordinate transformations, 
$$S= - {{m^2_{Pl}}\over{16 \pi}} \int {R \sqrt {-g} d^4x}+ \int {L \sqrt {-g} d^4 x}\  + $$
$$\ \ \ \ \ \ + {{A}\over{4!}} \int {\phi \ w^{\mu \nu \alpha \beta}_{\ \ ;\lambda}  w_{\mu \nu \alpha \beta}\ ^{;\lambda} \ d^4 x}
+ B \int {\phi (x) d^4 x} ,\eqno(3)$$
where $m_{Pl}$ and $L$ are the Planck mass and the Lagrangian density of the matter fields, respectively.  $A$ and $B$ are  real parameters of the theory with dimensions $(\mbox{mass})^2$ and $(\mbox{mass})^4$, respectively.  The portion of the
action in Eq.(3) pertaining to the four-form is by no means unique. For instance, we could add a term $\propto  \int {R \phi d^4x}$ to the above action, but at present  we restrict ourselves to gravitational minimal coupling. Later we
shall discuss a gravitational Chern-Simons like term involving $\phi $.

Since, $\tilde {w}$
is determined completely by the scalar density $\phi $ in a (3+1)-dimensional manifold, we will often refer to it as $\phi $ from now on. We mention in  passing that we could have raised  the indices 
of $w_{\mu \nu \rho \sigma }$ using  $g_{\mu \nu}$ to get a totally 
antisymmetric contravariant tensor $\equiv W^{\mu \nu \rho \sigma}=({\phi\over{\sqrt{-g}}})^2 w^{\mu \nu \rho \sigma}$ which, however, 
 is different  from the p-vector components $w^{\mu \nu \rho \sigma}$. 

By extremizing  $S$ with respect to $g_{\mu \nu}$ and $\phi $, respectively, we obtain the following equations of  motion,
$$R_{\mu \nu} - {1\over{2}} g_{\mu \nu} R = {{8 \pi}\over{m^2_{Pl}}} [T_{\mu \nu} + \Theta _{\mu \nu}] ,\eqno(4)$$
$$ \psi ^{\ ; \alpha}_{\ ; \alpha} \equiv {1\over{\sqrt {-g}}} (\sqrt {-g} g^{\alpha \beta} \psi_{ , \alpha})_{ , \beta} = {1\over {2}} \bigg [ g^{\mu \nu} {{\psi_{,\mu } \psi_{,\nu}}\over{\psi}} + {B\over {A}} \psi \bigg ] ,\eqno(5)$$ 
where $\psi (x) $, a scalar field variable,  is defined to be $\psi \equiv {\phi \over {\sqrt{-g}}} $, while $T_{\mu \nu}$ is the standard matter energy-momentum tensor  and  $\Theta _{\mu \nu}$ is the energy-momentum tensor 
for $\phi $  given by,
$$  \Theta _{\mu \nu} = 2 A [{{\psi_{,\mu } \psi_{,\nu}}\over{\psi}} - g_{\mu \nu} \psi ^{\ ; \alpha}_{\ ; \alpha}] .\eqno(6)$$
When $\phi $ satisfies the equation of motion given by Eq.(5), its energy-momentum tensor (Eq.(6)) takes the form,
$$  \Theta _{\mu \nu} = 2 A \bigg [{{\psi_{,\mu } \psi_{,\nu}}\over{\psi}} - {{g_{\mu \nu}}\over {2}} \bigg ( g^{\alpha \beta} {{\psi_{, \alpha } \psi_{,\beta}}\over{\psi}} + {B\over {A}} \psi \bigg )  \bigg] .\eqno(7)$$

The action for $\tilde {w} $ takes a simpler shape if one uses the variable $\psi $ instead of $\phi $ in Eq.(3),
$$S_\phi = \int {[{{A}\over{\psi}}  g^{\mu \nu} \psi_{, \mu} \psi_{, \nu} + B \psi ] \sqrt{-g} d^4 x} .\eqno(8)$$
 A point to be emphasized here is that although $g_{\mu \nu}$ and $\phi $ are mutually independent, the variable $\psi $ depends on both, so that when $g_{\mu \nu}\rightarrow $ $g_{\mu \nu} + \delta g_{\mu \nu}$,
 it induces $\psi \rightarrow $ $\psi + \delta \psi$, where $\delta \psi = {1\over{2}}\psi g_{\mu \nu} \delta g^{\mu \nu}$. Ofcourse, Eq.(4) then follows from $\delta S/ \delta g^{\mu \nu}=0$.

We now proceed to get a handle on the unknown  parameters $A$ and $B$ by invoking the current cosmological scenario.  For a flat Friedmann-Robertson-Walker (FRW) model,  Eq.(4) reduces to,
$${{\dot {a}}^2\over{a^2}}={{8 \pi}\over{3 m^2_{Pl}}}[T^0_{\ 0} + \Theta ^0_{\ 0}],\eqno(9)$$
and,
$$ 2 {{\ddot {a}}\over {a}} + {{\dot {a}}^2\over{a^2}}={{8 \pi}\over{m^2_{Pl}}}[T^1_{\ 1} + \Theta ^1_{\ 1}],\eqno(10)$$
where $T^0_{\ 0}=\epsilon$ and $\Theta^0_{\ 0}=A {{{\dot {\psi}}^2}\over{\psi}} - B \psi = \epsilon_\phi$ are the energy densities for matter and $\phi $, respectively, while 
$-T^1_{\ 1}=  p$ and $-\Theta ^1_{\ 1}=  A {{{\dot {\psi}}^2}\over{\psi}} + B \psi  =  p_\phi$ are the pressures for matter and the four-form, respectively.  For a flat FRW model, the equation of motion for $\phi $
given by Eq.(5) assumes the form,
$$ \ddot {\psi} + 3 {\dot {a}\over {a}} \dot {\psi} + {B_0\over {2}} \psi - {1\over{2}} {{\dot {\psi}}^2\over {\psi}}=0 ,\eqno(11)$$
where $B_0 \equiv -{B\over{A}}$. Eq.(11) can be turned into a second order linear differential equation by setting $\psi  = f ^2 $ so that,
$$ \ddot {f} + 3 {\dot {a}\over {a}} \dot {f} + {B_0\over {4}} f =0 .\eqno(12)$$
In terms of $f$, the energy density and pressure for $\phi $ are given by,
$$\epsilon _\phi = 4 A \dot {f}^2 - B f^2 = 4 A \bigg [ \dot {f}^2 + {B_0 \over {4} }f^2 \bigg ] \eqno(13)$$
and,
$$p _\phi = 4 A \dot {f}^2 + B f^2 = 4 A \bigg [ \dot {f}^2 - {B_0 \over {4} }f^2 \bigg ] ,\eqno(14)$$
respectively, and hence the expression for the equation of state parameter is simply,
$$w_\phi \equiv {{p_\phi}\over {\epsilon _ \phi}}= -(1 - {{4 {\dot f}^2}\over {B_0 f^2}})/(1 + {{4 {\dot f}^2}\over {B_0 f^2}}) .\eqno(15)$$

It is obvious from the above equation  that if $  {\dot f}^2 $ happens to be sufficiently small over a long stretch of time,  the value of $w_\phi $ is 
 $ \approx -1$, a required condition for an accelerated expansion of the 
universe. This raises the possibility of the scalar-density being
a source of DE.  According to Eqs.(13) and (14), $A > 0$ and $B < 0$ make $f$ mimic a quintessence field yielding $w_\phi > -1$, while $A < 0$
 and $B < 0$ reproduce the case of a phantom field with $w_\phi < -1$\cite{Ratra88, Wett, Cald, Zlat, Cald1}.  In the present work, we assume that  the kinetic energy and the mass terms 
for $f $ are positive definite so that $A > 0$ and $B < 0$, implying $B_0 > 0$. Bearing these points in mind, we study the
 dynamics  of $f$ after the universe has ceased to be radiation-dominated.
 
 Particles with high Lorentz gamma factors play an insignificant role in the evolution of the scale factor
 for redshifts less than $\approx 10^3$\cite{Wein72}. For such later epochs, if $\epsilon _\phi $ is negligible compared to the energy-density of 
 non-relativistic matter, universe is matter-dominated with scale factor  $ a \propto t^{2/3}$. This in turn implies  $3 {{\dot {a}}\over{a}} ={2\over {t}}$, leading to an exact solution of Eq.(12) 
 during the matter-dominated era,
 $$ f(t) = {{K_0}\over {t}} \sin ({{\sqrt{B_0} t}\over {2}} + K_1) ,\eqno(16)$$
 where $K_0$ and $K_1$ are constants of integration. 
 
 However, from times around the epoch of matter-$\phi $ equality $t_\phi \equiv 0.655 (2 / 3H_0)$, 
 the four-form energy density  $\epsilon _\phi $, far from being insignificant,  plays a major role in the dynamics of
   $a(t)$, since the density parameter $\Omega_{\phi 0}$ today is $\approx 0.7$ \cite{ Knop, Riess, Kowal}. Therefore,  Eqs.(9)-(11)  need to be solved in a self-consistent manner, with 
  the exact solution (Eq.(16)) acting as a rough guide.  In order to solve these coupled set of differential equations, we adopt the procedure of Dutta and Scherrer (2008). 
 
 Introducing a new function $g$ through \cite{Dutta08},
 $$f(t)= a^{-3/2} g(t)  \eqno(17)$$
  and substituting it in Eq.(12), we arrive at the following equation after using Eq.(10),
  $$ \ddot {g} +\bigg [{B_0\over {4}} + {{6 \pi p_{\phi}}\over {m^2_{Pl}}} \bigg ]g =0 .\eqno(18)$$
  In deriving the above equation, we have set the matter-pressure $p$ to zero in Eq.(10), as we are dealing with a `non-relativistic matter+ $\phi $' -dominated phase.
  
  During the early part of matter-dominated phase, $f$ is expected to trace the solution given by Eq. (16), displaying an oscillatory behaviour.  With a suitable choice of the phase $K_1$, one can arrange
  $f$ to reach a local maximum around the epoch $t_\phi$, and hover there till the present epoch.
  This essentially amounts to making a `slow-rolling' assumption,
  $${{4 {\dot f}^2}\over {B_0 f^2}} \ll 1 ,\eqno(19)$$
  and, hence when it is combined with Eq.(15), the outcome is,
  $$p_\phi \approx - \epsilon _\phi \eqno(20)$$
  
  Slow roll of $f$ (Eq.(19)) leads to $f$ and $\epsilon _\phi$ being roughly constant.  Then, from Eq.(9), with matter energy density going as $a^{-3}$, we have the standard solution,
$$a(t) = a_0 \bigg ({{\Omega_{m 0}}\over{\Omega_{\phi 0}}}\bigg )^{1/3} \sinh^{2\over{3}}\bigg ({3\over{2}}H_0 \sqrt{\Omega_{\phi 0}} t \bigg ),\eqno(21)$$
 with,
 $$\Omega_{\phi 0}\equiv {{8 \pi}\over {3 m^2_{Pl} H^2_0}} \epsilon_{\phi 0}\approx {{8 \pi A B_0 f^2}\over {3 m^2_{Pl} H^2_0}} ,\eqno(22)$$
 and $\Omega_{m 0} + \Omega_{\phi 0}=1$, for a flat FRW model.
  
 Making use of Eqs.(20) and (22) in Eq.(18), one obtains,
 $$ \ddot {g} + {1\over{4}} (B_0 -  B_1) g \approx 0 .\eqno(23)$$
 where $B_1 \equiv 9 H^2_0 \Omega_{\phi _0}$. 
 The preceding differential equation can be trivially solved. Defining $A_0 \equiv a^{-3/2}_0 \sqrt {{{\Omega_{\phi 0}}\over{\Omega_{m 0}}}}$, and combining Eqs. (17) with (21), one arrives at,
 $$f(t)= A_0 {{A_1\sin (\Delta . t + A_2)}\over{\sinh ({{\sqrt {B_1} t}\over{2}})}} , \ \ \ \ \ \ \mbox{for}  \ \  B_0 > B_1 \eqno(24a)$$
 $$\ \ \ \ \ \ = A_0 {{A_3\sinh (\gamma t) + A_4 \cosh (\gamma t)}\over{\sinh ({{\sqrt {B_1} t}\over{2}})}} , \ \ \mbox{for} \ B_0 < B_1 ,\eqno(24b)$$
where $\Delta \equiv {1\over{2}}\sqrt {B_0 - B_1}$, $\gamma \equiv {1\over{2}}\sqrt {B_1 - B_0}$ and $A_1,..., A_4$ are constants of integration. Comparison of Eqs.(24a) and (24b) with Eq.(16) along with the 
asymptotic behaviour of $f$ as $t \rightarrow \infty $, leads us to  the conclusion that the solution described by Eq.(24a) is physically more meaningful, imposing the restriction $B_0 > B_1$.  Having
narrowed the choice for $B_0$,
we turn our attention to the self-consistency of Eq.(24a) {\it {vis-a-vis}} the condition given by Eq.(19).

To study `slow rolling', we use Eq.(24a) to get the slope of $f$ so that,
$${{2 \dot {f}}\over{\sqrt{B_0} f}}= \sqrt {1-{1\over{r}}} \cot \bigg ({\sqrt{(r-1) B_1}\over{2}} t + A_2 \bigg ) -  {{\coth ({\sqrt{B_1}\over{2}} t )}\over{r^{1/2}}} , \eqno(25)$$
with $r \equiv {B_0\over {B_1}}$. If
 $A_2=0$ and $r=1+\epsilon^2 $, we have  from the above equation, 
 $${{2 \dot {f}}\over{\sqrt{B_0} f}} \approx ({\sqrt{B_1}\over{2}} t)^{-1} -\coth ({\sqrt{B_1}\over{2}} t), \eqno(26) $$ 
 for  $\epsilon ^2 $ infinitesmally small. Then, for $\epsilon \leq 10^{-1}$, as ${\sqrt{B_1}\over{2}} t$ goes from 0.1 to 1 we find that 
${{4 {\dot f}^2}\over {B_0 f^2}}$  varies from $10^{-3}$ to 0.1 satisfying thereby the condition given in Eq.(19). This is also  evident from Eq.(24a) since for $A_2=0$ and  for very small values of
$\Delta $ (i.e. $B_0 \approx B_1$), $f$ is
effectively flat over a long period of time. We therefore conclude that if the parameters $A$ and $B$ appearing in  Eq.(3) are such that  $- B$ is marginally in excess of
 $ \approx  9 A  H^2_0 \Omega_{\phi _0}$, the solution provided by Eq.(24a) is reasonably good, with the upshot being that $w_\phi \approx -1$ during $t=(0.18 -1.8) (2 / 3H_0)$. Although a specific range of values for the parameters 
 does reproduce the conditions necessary for the observed acceleration of the expansion factor,  the refrain of fine-tuning 
  haunts  this scenario too.
  
 It is interesting to observe that a Chern-Simons (CS) like term mirroring the coupling between electromagnetic field and the four-form $\tilde{w} $ arises very naturally in our model,
  \begin{eqnarray*}
S_{CS}= \mbox{J}\int {w^{\mu \nu \alpha \beta} F_{\mu \nu} A_\alpha \phi_{;\beta} \ d^4 x} \\
\ \ \ =\mbox {J} \int {\epsilon^{\mu \nu \alpha \beta} F_{\mu \nu} A_\alpha (\ln \psi)_{,\beta} \ d^4 x},
\ \ \ \ \ \ \ \ \ \ \ \ \ \ \ \ \ \ \ \ \ \ \ \ \ \ \ \ \ (27)
\end{eqnarray*}
where J is a  dimensionless constant, $F_{\mu \nu} = A_{\nu, \mu } - A_{\mu , \nu}$ and,
$$\phi_{; \beta} = \phi_{,\beta} - \Gamma^\alpha_{\alpha \beta}\phi  \eqno(28) $$ 
since $\phi $ is a
scalar-density of weight +1.
It is straightforward to establish that the above action is invariant under diffeomorphisms as well as gauge transformations, $A_\mu \rightarrow A_\mu + \partial_\mu \chi $, modulo boundary terms.
The outcome of adding $S_{CS}$ (Eq.(27)) to the standard action $S_{EM}$ for an electromagnetic field 
 interacting with charge particles in the presence of gravitation  is that, upon extremizing the full action $S_{EM} + S_{CS}$ with respect to $A_\mu$, one 
arrives at the following modified Einstein-Maxwell equation,
 $$F^{\alpha \beta}_{\ \ ; \beta}= -4 \pi j^\alpha + 8\pi \mbox{J}\  w^{\mu \nu \alpha \beta} F_{\mu \nu} \psi_{,\beta} \eqno(29)$$
 with $j^\alpha$ being the 4-current density associated with charge particles. 
 Applying the covariant derivative to the above equation with respect to $x^\alpha $ leads to the usual continuity equation implying conservation of electric charge,
 $$(\sqrt {-g} j^\alpha)_{, \alpha}=0   .\eqno(30)$$
 Eq.(29) opens up the possibility of  constraining the CS parameter J when used in conjunction with WMAP and other astrophysical  data. This requires further examination, and will
  form the subject of  a separate paper.

The CS-like analysis described above is similar to the one carried out by Jackiw and Pi (2003), except that  they employed an external fixed
 four-vector $v_{\mu}$ instead of a dynamical $\phi _{;\mu}$, entailing  violation of Lorentz invariance in their model \cite {Jackiw}.  In our case, both general as well as
 Lorentz covariance is maintained all through.  Inspired by the seminal work of  Jackiw and Pi, we undertake the exercise of studying the following gravitational CS term that involves $\tilde {w} $ and the Christoffel symbols, 
  \begin{eqnarray*}
S_{GCS}= \mbox{H}\int {w^{\mu \nu \alpha \beta} [\Gamma^\sigma_{\nu \tau} \partial _\alpha \Gamma^\tau_{\beta \sigma} + {2\over{3}} \Gamma^\sigma_{\nu \tau}\Gamma^\tau_{\alpha \eta}\Gamma^\eta_{\beta \sigma}]\phi_{;\mu} \ d^4 x} \\
\ \ \ =\mbox {H} \int {\epsilon^{\mu \nu \alpha \beta} [ \Gamma^\sigma_{\nu \tau} \partial _\alpha \Gamma^\tau_{\beta \sigma} + {2\over{3}} \Gamma^\sigma_{\nu \tau}\Gamma^\tau_{\alpha \eta}\Gamma^\eta_{\beta \sigma}](\ln \psi)_{,\mu} \ d^4 x}\ \ \ ,
\ \ \ \ \ \ \ \ \ \ \ \ \ \ \ \ \ \ (31)
\end{eqnarray*}
H being a dimensionless constant. After integrating by parts once, Eq.(31) can be expressed in terms of the Riemann tensor as,
$$S_{GCS}= -{\mbox{H}\over{2}} \int {\ln \psi *R R d^4 x}, \eqno(32)$$
where,
$$*R R \equiv {1\over{2}} \epsilon^{\mu \nu \alpha \beta}  R^\tau_{\ \sigma \alpha \beta}  R^\sigma_{\ \tau \mu \nu} = 8 \bigg [ R^{\tau \sigma}_ {\ \  01}  R_{\sigma  \tau  23} + R^{\tau \sigma}_ {\ \  12}  R_{\sigma  \tau  03}
+  R^{\tau \sigma}_ {\ \  13}  R_{\sigma  \tau  20}\bigg ]\eqno(33a)$$
and,
$$*R^{\tau \rho \mu \nu} \equiv {1\over{2}} \epsilon^{\mu \nu \alpha \beta} R^{\tau \rho}_ {\ \ \alpha \beta} \eqno(33b)$$
with,
$$R^\tau_{\ \sigma \alpha \beta} = \partial_\alpha \Gamma^\tau_{\sigma \beta} - \partial_\beta \Gamma^\tau_{\sigma \alpha} + \Gamma^\tau_{\alpha \eta} \Gamma^\eta_{\beta \sigma} - \Gamma^\tau_{\beta \eta} 
\Gamma^\eta_{\alpha \sigma} $$
and $R_{\alpha \beta} = R^\tau _{\ \alpha \tau \beta}$.

From Eq.(32) we find that  $\ln \psi $ acts like the parameter $\theta $ of   Jackiw and Pi 's paper in which the external vector $v_{\mu}= \theta_{,\mu}$. Adding $S_{CS} + S_{GCS}$ to $S$ of Eq.(3) and then extremizing
the total action with respect to $\phi $ and $g_{\mu \nu}$
leads to,
 $$ \psi ^{\ ; \alpha}_{\ ; \alpha}  = {1\over {2}} \bigg [ g^{\mu \nu} {{\psi_{,\mu } \psi_{,\nu}}\over{\psi}} + {B\over {A}} \psi  + {\mbox{J}\over{ 2 A}}\psi \ w^{\mu \nu \alpha \beta} F_{\mu \nu} F_{\alpha \beta}
-  {\mbox{H}\over{ 4 A}}\psi \  w^{\mu \nu \alpha \beta}   R^\tau_{\ \sigma \alpha \beta}  R^\sigma_{\ \tau \mu \nu} \bigg ] ,\eqno(34)$$ 
$$R_{\mu \nu} - {1\over{2}} g_{\mu \nu} R = {{8 \pi}\over{m^2_{Pl}}} [T_{\mu \nu} + \Theta _{\mu \nu} + C_{\mu \nu}] ,\eqno(35)$$
where the modified Cotton tensor $C^{\mu \nu}$ is defined as,
$$C^{\mu \nu} \equiv -2{\mbox{H}\over{\sqrt{-g}}} \bigg [{1\over{4}} *R R g^{\mu \nu} - (\ln \psi)_{; \alpha ; \beta} \bigg ( *R^{\beta \mu \alpha \nu} + *R^{\beta \nu \alpha \mu} \bigg ) + (\ln \psi)_{, \alpha}
\bigg (\epsilon^{\alpha \mu \sigma \tau} R^\nu _{\ \sigma ; \tau} + \epsilon^{\alpha \nu \sigma \tau} R^\mu _{\ \sigma ; \tau} \bigg ) \bigg ] . \eqno(36)$$
The first term in the RHS of Eq.(36) is new and is not present in the expression for the Cotton tensor  as delineated by Jackiw and Pi. Here it appears because under an infinitesmal variation
$g_{\mu \nu} \rightarrow$ $g_{\mu \nu} + \delta g_{\mu \nu}$, the change in $(\ln \psi )_{,\mu}$ occurring in Eq.(31) is given by,
$$\delta (\ln \psi)_{,\mu} = \delta ( \phi_{; \mu} / \phi)= - \delta \Gamma^\alpha_{\alpha \mu}= -{1\over{2}} \delta ( g^{\alpha \beta} g_{\alpha \beta , \mu}) \eqno(37)$$
which follows after making use of Eq.(28). 

In the absence electromagnetic field, when the equation of motion for the dynamical four-form given by eq.(34) is substituted in eq.(6), its energy-momentum tensor  in the
 presence of gravitational CS-term takes the form,
$$\Theta _{\mu \nu}= 2 A \bigg [{{\psi_{,\mu } \psi_{,\nu}}\over{\psi}} - {{g_{\mu \nu}}\over {2}} \bigg ( g^{\alpha \beta} {{\psi_{, \alpha } \psi_{,\beta}}\over{\psi}} +
 {B\over {A}} \psi \bigg )  \bigg] + {H\over{2\sqrt{-g}}} *R R g_{\mu \nu} \eqno(38)$$
 so that when eq.(38) is substituted in eq.(35) the first term in the RHS of eq.(36) cancels with the last term in the RHS of eq.(38). In other words, the
 term $*R R g_{\mu \nu}$ does not contribute to the Einstein equations, and it is the standard Cotton tensor \cite {Jackiw} along with $\Theta _{\mu \nu}$ given
 by eq.(7) that appear in the RHS of eq.(35).

The natural question to ask is with the entry of $S_{GCS}$ can the dynamics of $\phi $, now governed  by Eqs.(34) and (35), still lead to an accelerating universe? We show that the answer is in the affirmative. In the 
context of a flat FRW model, all the non-zero components of the Riemann tensor are obtained from,
\begin{eqnarray*}
R_{0101}= a \ddot{a} \ \ \  R_{0202}= r^2  a \ddot{a} \ \ \ R_{0303}= (r \sin {\theta})^2 a \ddot{a} \\ 
R_{1212}= - r^2 (a \dot {a})^2 \ \ \ R_{1313}=\sin ^2{\theta} R_{1212} \ \ \ R_{2323}=r^2 R_{1313} \ \ \ \ \ \ \ \ \ \ (39)
\end{eqnarray*}
Using Eq.(39) in Eq.(33a), one readily verifies that $*R R=0$. Hence, in the absence of electromagnetic fields, the equation of motion for $\phi$ given by eq.(34) 
 simply reduces to Eq.(5) as the last term in the RHS of Eq.(34) is proportional to $*RR$. Since in the case of FRW models, only the diagonal components of Ricci tensor are non-zero 
with $R^1_{\ 1}$ = $R^2_{\ 2}$ = $R^3_{\ 3}$ and that  they depend on time alone, we find after tedious algebra involving Eqs.(39) and (33b) that even the second and third terms in Eq.(36) vanish.  This
explicitly demonstrates that the modified Cotton tensor is zero for a flat FRW universe implying that the inclusion of $S_{GCS}$ does not alter our solution for $f(t)$ given by Eq.(24a) that is responsible
for an accelerating universe, so long as `slow rolling' condition is met. We could have reached this conclusion straightaway without labourious calculations from the fact that $*R R=0$ implies $S_{GCS}=0$ from
Eq.(32).

 Apart from the possibility of adding `Chern-Simons' like terms described above,  and a natural coupling that may arise between $\tilde {w}$ and branes in string/M-theories \cite{Bousso, Feng, Dvali, Garr, Kall, Wohl, Garr1, Jarv}, there may also 
 be a differential geometric significance of the scalar-density $\phi $,
 in the sense that  it can be used  to
 generate  an antiderivation on antisymmetric tensor-densities  of arbitrary weights. If $\tilde {\alpha} $ is a p-form density with weight $w$ such that its components  
 $\alpha _{\nu_1 \nu_2 .. \nu_p} $ transform to $J^{-w} \alpha _{\nu_1 \nu_2 .. \nu_p }$, under a general coordinate transformation, $J$ being the Jacobian, then we define a generalized exterior derivative $\tilde {d}_w$
in the following manner,
\begin{eqnarray*}
\tilde {d}_w \tilde {\alpha} \equiv {1\over{p!}}\partial_ \mu \alpha _{\nu_1 \nu_2 .. \nu_p } \tilde {d} x^\mu \wedge \tilde {d} x^{\nu _1} \wedge ... \wedge \tilde {d} x^{\nu _p} -\\
- w \partial_\mu (\ln \phi) \tilde {d} x^\mu \wedge  \tilde {\alpha} 
= \tilde {d} \tilde {\alpha}  - w  \tilde {d}\ (\ln \phi) \wedge  \tilde {\alpha} .
\end{eqnarray*}
It is easy to see that $\tilde {d}_w \tilde {\alpha} $ is a (p+1)-form density of weight $w$. 

If $\chi _1 $ and $\chi_2 $   are scalar-densities of weights $w_1$ and $w_2 $, respectively, then the above equation leads to,
$$\tilde {d}_w \chi_i = \partial_\mu  \chi_i  \tilde {d} x^\mu  - w _i \chi_i \partial_\mu (\ln \phi) \tilde {d} x^\mu  
, \ \ \ i=1,2 \eqno(39a)$$
 being one-form densities and, furthermore, one can show that,
$$\tilde {d}_w (\chi_1 \tilde {d}_w \chi_2) =  \tilde {d}_w \chi_1 \wedge \tilde {d}_w \chi_2 . \eqno(39b)$$
The generalized exterior derivative also satisfies (a) $\tilde {d}_w \tilde {d}_w =0 $ and  (b) $\tilde {d}_w (\tilde {\alpha} \wedge \tilde {\beta}) 
= \tilde {d}_w \tilde {\alpha} \wedge \tilde {\beta} + (-1)^p \tilde {\alpha } \wedge \tilde {d}_w \tilde {\beta} $, where $ \tilde {\alpha} $ and $\tilde {\beta} $ are p- and q-form densities of weights $w_1 $ and $w_2 $, respectively.
 These properties are sufficient to qualify $\tilde {d}_w $ to the role of a well-defined antiderivation on differential form-densities \cite {Muku}.  For instance, from Eq.(39a) it follows that,
$$\tilde {d}_w \sqrt {-g}  = - \sqrt {-g} \ \tilde {d} \ln ({\phi \over {\sqrt {-g}}}) ,  \eqno(40)$$
since $\sqrt {-g} $ is a scalar-density of weight +1. There are other physically meaningful antisymmetric tensor-densities, e.g. dual of $F^{\mu \nu}$, on which we may apply $\tilde {d}_w $\cite {Berg, Schro}. 
 It is interesting to note that  $\tilde {d}_w \phi = 0$.  This is 
 analogous to the vanishing of $g_{\mu \nu ; \lambda} $. 
 
 To summarize in a nutshell, what we have demonstrated in this paper is that if Einstein's geometrical theory of gravitation is extended by including a new degree of freedom  $\tilde {w} $ that is independent of the 
 metric $g_{\mu \nu}$,  there exists a long band of allowed region in the parameter-plane constituted by $A$ and $B$, such that the dynamics of $\phi $ 
  does give rise to an accelerating universe in the context of a flat FRW model. We concede here that even this model cannot shake off the fine-tuning problem that plagues other DE models, as it too relies on the `slow rolling'
  condition to attain $w_{\phi}\approx -1$. 
  
  A Chern-Simons like coupling between electromagnetic fields and $\phi $ comes about naturally, causing  a modification of Einstein-Maxwell equation. What ensues from a  gravitational Chern-Simons like term is that 
  a modified Cotton tensor appears in the Einstein equation, although this has no effect on the dynamics in a flat FRW universe. 
  The scalar-density associated with  $\tilde {w} $ leads to a well-defined exterior derivative that turns a differential p-form density into 
   a (p+1)- form 
  density of same
 weight.  Since the notion of an n-form and exterior derivative in a differential manifold does not require either an affine connection or a metric,  further studies are required
 to investigate the role of $\tilde {w} $ in situations where metric is ill-defined and in the possibility 
 of  its causing transitions in manifold-orientability.
\subsection{}
\subsubsection{}

\begin{acknowledgments}
I thank N. Mukunda and Joseph Samuel for drawing my attention to some key references. I also acknowledge  technical assistance extended to me by
Ujjwal Dasgupta. This research has made use of NASA's Astrophysics Data System.
\end{acknowledgments}

\bibliographystyle{apsrev}
\bibliography{fourform}

\begin{thebibliography}{43}
\expandafter\ifx\csname natexlab\endcsname\relax\def\natexlab#1{#1}\fi
\expandafter\ifx\csname bibnamefont\endcsname\relax
  \def\bibnamefont#1{#1}\fi
\expandafter\ifx\csname bibfnamefont\endcsname\relax
  \def\bibfnamefont#1{#1}\fi
\expandafter\ifx\csname citenamefont\endcsname\relax
  \def\citenamefont#1{#1}\fi
\expandafter\ifx\csname url\endcsname\relax
  \def\url#1{\texttt{#1}}\fi
\expandafter\ifx\csname urlprefix\endcsname\relax\def\urlprefix{URL }\fi
\providecommand{\bibinfo}[2]{#2}
\providecommand{\eprint}[2][]{\url{#2}}

\bibitem[{\citenamefont{Perlmutter et~al.}(1997)\citenamefont{Perlmutter, Gabi,
  Goldhaber, Groom, Hook, Kim, Kim, Lee, Pennypacker, Small et~al.}}]{Perl}
\bibinfo{author}{\bibfnamefont{S.}~\bibnamefont{Perlmutter}},
  \bibinfo{author}{\bibfnamefont{S.}~\bibnamefont{Gabi}},
  \bibinfo{author}{\bibfnamefont{G.}~\bibnamefont{Goldhaber}},
  \bibinfo{author}{\bibfnamefont{D.~E.} \bibnamefont{Groom}},
  \bibinfo{author}{\bibfnamefont{I.~M.} \bibnamefont{Hook}},
  \bibinfo{author}{\bibfnamefont{A.~G.} \bibnamefont{Kim}},
  \bibinfo{author}{\bibfnamefont{M.~Y.} \bibnamefont{Kim}},
  \bibinfo{author}{\bibfnamefont{J.~C.} \bibnamefont{Lee}},
  \bibinfo{author}{\bibfnamefont{C.~R.} \bibnamefont{Pennypacker}},
  \bibinfo{author}{\bibfnamefont{I.~A.} \bibnamefont{Small}},
  \bibnamefont{et~al.}, \bibinfo{journal}{Ap.J.}
  \textbf{\bibinfo{volume}{483}}, \bibinfo{pages}{565} (\bibinfo{year}{1997}).

\bibitem[{\citenamefont{Schmidt et~al.}(1998)\citenamefont{Schmidt, Suntzeff,
  Phillips, Schommer, Clocchiatti, Kirshner, Garnavich, Challis, Leibundgut,
  Spyromilio et~al.}}]{Schmi}
\bibinfo{author}{\bibfnamefont{B.~P.} \bibnamefont{Schmidt}},
  \bibinfo{author}{\bibfnamefont{N.~B.} \bibnamefont{Suntzeff}},
  \bibinfo{author}{\bibfnamefont{M.~M.} \bibnamefont{Phillips}},
  \bibinfo{author}{\bibfnamefont{R.~A.} \bibnamefont{Schommer}},
  \bibinfo{author}{\bibfnamefont{A.}~\bibnamefont{Clocchiatti}},
  \bibinfo{author}{\bibfnamefont{R.~P.} \bibnamefont{Kirshner}},
  \bibinfo{author}{\bibfnamefont{P.}~\bibnamefont{Garnavich}},
  \bibinfo{author}{\bibfnamefont{P.}~\bibnamefont{Challis}},
  \bibinfo{author}{\bibfnamefont{B.}~\bibnamefont{Leibundgut}},
  \bibinfo{author}{\bibfnamefont{J.}~\bibnamefont{Spyromilio}},
  \bibnamefont{et~al.}, \bibinfo{journal}{Ap. J.}
  \textbf{\bibinfo{volume}{507}}, \bibinfo{pages}{46} (\bibinfo{year}{1998}).

\bibitem[{\citenamefont{Riess et~al.}(1998)\citenamefont{Riess, Filippenko,
  Challis, Clocchiattia, Diercks, Garnavich, Gilliland, Hogan, Jha, Kirshner
  et~al.}}]{Riess98}
\bibinfo{author}{\bibfnamefont{A.~G.} \bibnamefont{Riess}},
  \bibinfo{author}{\bibfnamefont{A.~V.} \bibnamefont{Filippenko}},
  \bibinfo{author}{\bibfnamefont{P.}~\bibnamefont{Challis}},
  \bibinfo{author}{\bibfnamefont{A.}~\bibnamefont{Clocchiattia}},
  \bibinfo{author}{\bibfnamefont{A.}~\bibnamefont{Diercks}},
  \bibinfo{author}{\bibfnamefont{P.~M.} \bibnamefont{Garnavich}},
  \bibinfo{author}{\bibfnamefont{R.~L.} \bibnamefont{Gilliland}},
  \bibinfo{author}{\bibfnamefont{C.~J.} \bibnamefont{Hogan}},
  \bibinfo{author}{\bibfnamefont{S.}~\bibnamefont{Jha}},
  \bibinfo{author}{\bibfnamefont{R.~P.} \bibnamefont{Kirshner}},
  \bibnamefont{et~al.}, \bibinfo{journal}{A. J.}
  \textbf{\bibinfo{volume}{116}}, \bibinfo{pages}{1009} (\bibinfo{year}{1998}).

\bibitem[{\citenamefont{Kowalski et~al.}(2008)\citenamefont{Kowalski, D.Rubin,
  G.Aldering, R.J.Agostinho, A.Amadon, R.Amanullah, C.Balland, Barbary,
  G.Blanc, P.J.Challis et~al.}}]{Kowal}
\bibinfo{author}{\bibfnamefont{M.}~\bibnamefont{Kowalski}},
  \bibinfo{author}{\bibnamefont{D.Rubin}},
  \bibinfo{author}{\bibnamefont{G.Aldering}},
  \bibinfo{author}{\bibnamefont{R.J.Agostinho}},
  \bibinfo{author}{\bibnamefont{A.Amadon}},
  \bibinfo{author}{\bibnamefont{R.Amanullah}},
  \bibinfo{author}{\bibnamefont{C.Balland}},
  \bibinfo{author}{\bibfnamefont{K.}~\bibnamefont{Barbary}},
  \bibinfo{author}{\bibnamefont{G.Blanc}},
  \bibinfo{author}{\bibnamefont{P.J.Challis}}, \bibnamefont{et~al.},
  \bibinfo{journal}{Ap. J.} \textbf{\bibinfo{volume}{686}},
  \bibinfo{pages}{749} (\bibinfo{year}{2008}).

\bibitem[{\citenamefont{Weinberg}(1989)}]{Wein89}
\bibinfo{author}{\bibfnamefont{S.}~\bibnamefont{Weinberg}},
  \bibinfo{journal}{Rev. Mod. Phys.} \textbf{\bibinfo{volume}{61}},
  \bibinfo{pages}{1} (\bibinfo{year}{1989}).

\bibitem[{\citenamefont{Vilenkin}(2003)}]{Vilen}
\bibinfo{author}{\bibfnamefont{A.}~\bibnamefont{Vilenkin}}, in
  \emph{\bibinfo{booktitle}{The Dark Universe: Matter, Energy, and Gravity}},
  edited by \bibinfo{editor}{\bibfnamefont{M.}~\bibnamefont{Livio}}
  (\bibinfo{publisher}{Cambridge University Press},
  \bibinfo{address}{Cambridge}, \bibinfo{year}{2003}).

\bibitem[{\citenamefont{Peebles and Ratra}(1988)}]{Ratra88}
\bibinfo{author}{\bibfnamefont{P.~J.~E.} \bibnamefont{Peebles}}
  \bibnamefont{and} \bibinfo{author}{\bibfnamefont{B.}~\bibnamefont{Ratra}},
  \bibinfo{journal}{Ap. J. Lett.} \textbf{\bibinfo{volume}{325}},
  \bibinfo{pages}{L17} (\bibinfo{year}{1988}).

\bibitem[{\citenamefont{Wetterich}(1988)}]{Wett}
\bibinfo{author}{\bibfnamefont{C.}~\bibnamefont{Wetterich}},
  \bibinfo{journal}{Nucl. Phys. B} \textbf{\bibinfo{volume}{302}},
  \bibinfo{pages}{668} (\bibinfo{year}{1988}).

\bibitem[{\citenamefont{Caldwell et~al.}(1998)\citenamefont{Caldwell, Dave, and
  Steinhardt}}]{Cald}
\bibinfo{author}{\bibfnamefont{R.~R.} \bibnamefont{Caldwell}},
  \bibinfo{author}{\bibfnamefont{R.}~\bibnamefont{Dave}}, \bibnamefont{and}
  \bibinfo{author}{\bibfnamefont{P.~J.} \bibnamefont{Steinhardt}},
  \bibinfo{journal}{Phys. Rev.Lett.} \textbf{\bibinfo{volume}{80}},
  \bibinfo{pages}{1582} (\bibinfo{year}{1998}).

\bibitem[{\citenamefont{Zlatev et~al.}(1999)\citenamefont{Zlatev, Wang, and
  Steinhardt}}]{Zlat}
\bibinfo{author}{\bibfnamefont{I.}~\bibnamefont{Zlatev}},
  \bibinfo{author}{\bibfnamefont{L.}~\bibnamefont{Wang}}, \bibnamefont{and}
  \bibinfo{author}{\bibfnamefont{P.~J.} \bibnamefont{Steinhardt}},
  \bibinfo{journal}{Phys. Rev. Lett.} \textbf{\bibinfo{volume}{82}},
  \bibinfo{pages}{896} (\bibinfo{year}{1999}).

\bibitem[{\citenamefont{Dutta and Scherrer}(2008)}]{Dutta08}
\bibinfo{author}{\bibfnamefont{S.}~\bibnamefont{Dutta}} \bibnamefont{and}
  \bibinfo{author}{\bibfnamefont{R.~J.} \bibnamefont{Scherrer}},
  \bibinfo{journal}{Phys. Rev. D} \textbf{\bibinfo{volume}{78}},
  \bibinfo{pages}{123525} (\bibinfo{year}{2008}).

\bibitem[{\citenamefont{Caldwell}(2002)}]{Cald1}
\bibinfo{author}{\bibfnamefont{R.~R.} \bibnamefont{Caldwell}},
  \bibinfo{journal}{Phys. Lett. B} \textbf{\bibinfo{volume}{545}},
  \bibinfo{pages}{23} (\bibinfo{year}{2002}).

\bibitem[{\citenamefont{Aurilia et~al.}(1978)\citenamefont{Aurilia,
  Christodoulou, and Legovini}}]{Auri78}
\bibinfo{author}{\bibfnamefont{A.}~\bibnamefont{Aurilia}},
  \bibinfo{author}{\bibfnamefont{D.}~\bibnamefont{Christodoulou}},
  \bibnamefont{and} \bibinfo{author}{\bibfnamefont{F.}~\bibnamefont{Legovini}},
  \bibinfo{journal}{Phys. Lett. B} \textbf{\bibinfo{volume}{73}},
  \bibinfo{pages}{429} (\bibinfo{year}{1978}).

\bibitem[{\citenamefont{Duff and van Nieuwenhuizen}(1980)}]{Duff}
\bibinfo{author}{\bibfnamefont{M.~J.} \bibnamefont{Duff}} \bibnamefont{and}
  \bibinfo{author}{\bibfnamefont{P.}~\bibnamefont{van Nieuwenhuizen}},
  \bibinfo{journal}{Phys. Lett. B} \textbf{\bibinfo{volume}{94}},
  \bibinfo{pages}{179} (\bibinfo{year}{1980}).

\bibitem[{\citenamefont{Aurilia et~al.}(1980)\citenamefont{Aurilia, Nicolai,
  and Townsend}}]{Auri80}
\bibinfo{author}{\bibfnamefont{A.}~\bibnamefont{Aurilia}},
  \bibinfo{author}{\bibfnamefont{H.}~\bibnamefont{Nicolai}}, \bibnamefont{and}
  \bibinfo{author}{\bibfnamefont{P.~K.} \bibnamefont{Townsend}},
  \bibinfo{journal}{Nucl. Phys. B} \textbf{\bibinfo{volume}{176}},
  \bibinfo{pages}{509} (\bibinfo{year}{1980}).

\bibitem[{\citenamefont{Hawking}(1984)}]{Hawk}
\bibinfo{author}{\bibfnamefont{S.~W.} \bibnamefont{Hawking}},
  \bibinfo{journal}{Phys. Lett. B} \textbf{\bibinfo{volume}{134}},
  \bibinfo{pages}{403} (\bibinfo{year}{1984}).

\bibitem[{\citenamefont{Brown and Teitelboim}(1988)}]{Brown}
\bibinfo{author}{\bibfnamefont{J.~D.} \bibnamefont{Brown}} \bibnamefont{and}
  \bibinfo{author}{\bibfnamefont{C.}~\bibnamefont{Teitelboim}},
  \bibinfo{journal}{Nucl. Phys. B} \textbf{\bibinfo{volume}{297}},
  \bibinfo{pages}{787} (\bibinfo{year}{1988}).

\bibitem[{\citenamefont{Duff}(1989)}]{Duff1}
\bibinfo{author}{\bibfnamefont{M.~J.} \bibnamefont{Duff}},
  \bibinfo{journal}{Phys. Lett. B} \textbf{\bibinfo{volume}{226}},
  \bibinfo{pages}{36} (\bibinfo{year}{1989}).

\bibitem[{\citenamefont{Duncan and Jensen}(1990)}]{Duncan}
\bibinfo{author}{\bibfnamefont{M.~J.} \bibnamefont{Duncan}} \bibnamefont{and}
  \bibinfo{author}{\bibfnamefont{L.~G.} \bibnamefont{Jensen}},
  \bibinfo{journal}{Nucl. Phys. B} \textbf{\bibinfo{volume}{336}},
  \bibinfo{pages}{100} (\bibinfo{year}{1990}).

\bibitem[{\citenamefont{Turok and Hawking}(1998)}]{Turok}
\bibinfo{author}{\bibfnamefont{N.}~\bibnamefont{Turok}} \bibnamefont{and}
  \bibinfo{author}{\bibfnamefont{S.~W.} \bibnamefont{Hawking}},
  \bibinfo{journal}{Phys. Lett. B} \textbf{\bibinfo{volume}{432}},
  \bibinfo{pages}{271} (\bibinfo{year}{1998}).

\bibitem[{\citenamefont{Aurilia and Spallucci}(2004)}]{Auri04}
\bibinfo{author}{\bibfnamefont{A.}~\bibnamefont{Aurilia}} \bibnamefont{and}
  \bibinfo{author}{\bibfnamefont{E.}~\bibnamefont{Spallucci}},
  \bibinfo{journal}{Phys. Rev. D} \textbf{\bibinfo{volume}{69}},
  \bibinfo{pages}{105004} (\bibinfo{year}{2004}).

\bibitem[{\citenamefont{Wu}(2008)}]{Wu}
\bibinfo{author}{\bibfnamefont{Z.~C.} \bibnamefont{Wu}},
  \bibinfo{journal}{Phys. Lett. B} \textbf{\bibinfo{volume}{659}},
  \bibinfo{pages}{891} (\bibinfo{year}{2008}).

\bibitem[{\citenamefont{Klinkhamer and Volovik}(2008)}]{Klink}
\bibinfo{author}{\bibfnamefont{F.~R.} \bibnamefont{Klinkhamer}}
  \bibnamefont{and} \bibinfo{author}{\bibfnamefont{G.~E.}
  \bibnamefont{Volovik}}, \bibinfo{journal}{Phys. Rev. D}
  \textbf{\bibinfo{volume}{78}}, \bibinfo{pages}{063528}
  (\bibinfo{year}{2008}).

\bibitem[{\citenamefont{Bousso and Polchinski}(2000)}]{Bousso}
\bibinfo{author}{\bibfnamefont{R.}~\bibnamefont{Bousso}} \bibnamefont{and}
  \bibinfo{author}{\bibfnamefont{J.}~\bibnamefont{Polchinski}},
  \bibinfo{journal}{JHEP} \textbf{\bibinfo{volume}{0006}}, \bibinfo{pages}{006}
  (\bibinfo{year}{2000}).

\bibitem[{\citenamefont{Feng et~al.}(2001)\citenamefont{Feng, March-Russell,
  Sethi, and Wilczek}}]{Feng}
\bibinfo{author}{\bibfnamefont{J.~L.} \bibnamefont{Feng}},
  \bibinfo{author}{\bibfnamefont{J.}~\bibnamefont{March-Russell}},
  \bibinfo{author}{\bibfnamefont{S.}~\bibnamefont{Sethi}}, \bibnamefont{and}
  \bibinfo{author}{\bibfnamefont{F.}~\bibnamefont{Wilczek}},
  \bibinfo{journal}{Nucl. Phys. B} \textbf{\bibinfo{volume}{602}},
  \bibinfo{pages}{307} (\bibinfo{year}{2001}).

\bibitem[{\citenamefont{Dvali and Vilenkin}(2001)}]{Dvali}
\bibinfo{author}{\bibfnamefont{G.}~\bibnamefont{Dvali}} \bibnamefont{and}
  \bibinfo{author}{\bibfnamefont{A.}~\bibnamefont{Vilenkin}},
  \bibinfo{journal}{Phys. Rev. D} \textbf{\bibinfo{volume}{64}},
  \bibinfo{pages}{063509} (\bibinfo{year}{2001}).

\bibitem[{\citenamefont{Garriga and Vilenkin}(2003)}]{Garr}
\bibinfo{author}{\bibfnamefont{J.}~\bibnamefont{Garriga}} \bibnamefont{and}
  \bibinfo{author}{\bibfnamefont{A.}~\bibnamefont{Vilenkin}},
  \bibinfo{journal}{Phys. Rev. D} \textbf{\bibinfo{volume}{67}},
  \bibinfo{pages}{043503} (\bibinfo{year}{2003}).

\bibitem[{\citenamefont{Kallosh and Linde}(2003)}]{Kall}
\bibinfo{author}{\bibfnamefont{R.}~\bibnamefont{Kallosh}} \bibnamefont{and}
  \bibinfo{author}{\bibfnamefont{A.}~\bibnamefont{Linde}},
  \bibinfo{journal}{Phys. Rev. D} \textbf{\bibinfo{volume}{67}},
  \bibinfo{pages}{023510} (\bibinfo{year}{2003}).

\bibitem[{\citenamefont{Wohlfarth}(2003)}]{Wohl}
\bibinfo{author}{\bibfnamefont{M.~N.~R.} \bibnamefont{Wohlfarth}},
  \bibinfo{journal}{Phys. Lett. B} \textbf{\bibinfo{volume}{563}},
  \bibinfo{pages}{1} (\bibinfo{year}{2003}).

\bibitem[{\citenamefont{Garriga and Megevand}(2004)}]{Garr1}
\bibinfo{author}{\bibfnamefont{J.}~\bibnamefont{Garriga}} \bibnamefont{and}
  \bibinfo{author}{\bibfnamefont{A.}~\bibnamefont{Megevand}},
  \bibinfo{journal}{Phys. Rev. D} \textbf{\bibinfo{volume}{69}},
  \bibinfo{pages}{083510} (\bibinfo{year}{2004}).

\bibitem[{\citenamefont{J$\ddot{\mbox{a}}$rv
  et~al.}(2004)\citenamefont{J$\ddot{\mbox{a}}$rv, Mohaupt, and
  Saueressig}}]{Jarv}
\bibinfo{author}{\bibfnamefont{L.}~\bibnamefont{J$\ddot{\mbox{a}}$rv}},
  \bibinfo{author}{\bibfnamefont{T.}~\bibnamefont{Mohaupt}}, \bibnamefont{and}
  \bibinfo{author}{\bibfnamefont{F.}~\bibnamefont{Saueressig}},
  \bibinfo{journal}{JCAP} \textbf{\bibinfo{volume}{0408}}, \bibinfo{pages}{016}
  (\bibinfo{year}{2004}).

\bibitem[{\citenamefont{Graf}(2003)}]{Graf}
\bibinfo{author}{\bibfnamefont{W.}~\bibnamefont{Graf}}, \bibinfo{journal}{Phys.
  Rev. D} \textbf{\bibinfo{volume}{67}}, \bibinfo{pages}{024002}
  (\bibinfo{year}{2003}).

\bibitem[{\citenamefont{Graf}(2007)}]{Graf1}
\bibinfo{author}{\bibfnamefont{W.}~\bibnamefont{Graf}},
  \bibinfo{journal}{arXiv:gr-qc} \textbf{\bibinfo{volume}{0602054}}
  (\bibinfo{year}{2007}).

\bibitem[{\citenamefont{Guendelman and Kaganovich}(2008)}]{Guen}
\bibinfo{author}{\bibfnamefont{E.~I.} \bibnamefont{Guendelman}}
  \bibnamefont{and} \bibinfo{author}{\bibfnamefont{A.~B.}
  \bibnamefont{Kaganovich}}, \bibinfo{journal}{Class. Quant. Grav.}
  \textbf{\bibinfo{volume}{25}}, \bibinfo{pages}{235015}
  (\bibinfo{year}{2008}).

\bibitem[{\citenamefont{Koivisto et~al.}(2009)\citenamefont{Koivisto, Mota, and
  Pitrou}}]{Mota}
\bibinfo{author}{\bibfnamefont{T.~S.} \bibnamefont{Koivisto}},
  \bibinfo{author}{\bibfnamefont{D.~F.} \bibnamefont{Mota}}, \bibnamefont{and}
  \bibinfo{author}{\bibfnamefont{C.}~\bibnamefont{Pitrou}},
  \bibinfo{journal}{arXiv:astro-ph.CO} \textbf{\bibinfo{volume}{0903.4158}}
  (\bibinfo{year}{2009}).

\bibitem[{\citenamefont{Weinberg}(1972)}]{Wein72}
\bibinfo{author}{\bibfnamefont{S.}~\bibnamefont{Weinberg}},
  \emph{\bibinfo{title}{Gravitation and Cosmology}} (\bibinfo{publisher}{John
  Wiley and Sons}, \bibinfo{address}{New York}, \bibinfo{year}{1972}).

\bibitem[{\citenamefont{Schutz}(1980)}]{Schutz80}
\bibinfo{author}{\bibfnamefont{B.~F.} \bibnamefont{Schutz}},
  \emph{\bibinfo{title}{Geometrical methods of mathematical physics}}
  (\bibinfo{publisher}{Cambridge University Press},
  \bibinfo{address}{Cambridge}, \bibinfo{year}{1980}).

\bibitem[{\citenamefont{Knop et~al.}(2003)\citenamefont{Knop, Aldering,
  Amanullah, Astier, Blanc, Burns, Conley, Deustua, Doi, Ellis et~al.}}]{Knop}
\bibinfo{author}{\bibfnamefont{R.~A.} \bibnamefont{Knop}},
  \bibinfo{author}{\bibfnamefont{G.}~\bibnamefont{Aldering}},
  \bibinfo{author}{\bibfnamefont{R.}~\bibnamefont{Amanullah}},
  \bibinfo{author}{\bibfnamefont{P.}~\bibnamefont{Astier}},
  \bibinfo{author}{\bibfnamefont{G.}~\bibnamefont{Blanc}},
  \bibinfo{author}{\bibfnamefont{M.~S.} \bibnamefont{Burns}},
  \bibinfo{author}{\bibfnamefont{A.}~\bibnamefont{Conley}},
  \bibinfo{author}{\bibfnamefont{S.~E.} \bibnamefont{Deustua}},
  \bibinfo{author}{\bibfnamefont{M.}~\bibnamefont{Doi}},
  \bibinfo{author}{\bibfnamefont{R.}~\bibnamefont{Ellis}},
  \bibnamefont{et~al.}, \bibinfo{journal}{Ap. J.}
  \textbf{\bibinfo{volume}{598}}, \bibinfo{pages}{102} (\bibinfo{year}{2003}).

\bibitem[{\citenamefont{Riess et~al.}(2004)\citenamefont{Riess, Strolger,
  Tonry, Casertano, Ferguson, Mobasher, Challis, Filippenko, Jha, Li
  et~al.}}]{Riess}
\bibinfo{author}{\bibfnamefont{A.~G.} \bibnamefont{Riess}},
  \bibinfo{author}{\bibfnamefont{L.-G.} \bibnamefont{Strolger}},
  \bibinfo{author}{\bibfnamefont{J.}~\bibnamefont{Tonry}},
  \bibinfo{author}{\bibfnamefont{S.}~\bibnamefont{Casertano}},
  \bibinfo{author}{\bibfnamefont{H.~C.} \bibnamefont{Ferguson}},
  \bibinfo{author}{\bibfnamefont{B.}~\bibnamefont{Mobasher}},
  \bibinfo{author}{\bibfnamefont{P.}~\bibnamefont{Challis}},
  \bibinfo{author}{\bibfnamefont{A.~V.} \bibnamefont{Filippenko}},
  \bibinfo{author}{\bibfnamefont{S.}~\bibnamefont{Jha}},
  \bibinfo{author}{\bibfnamefont{W.}~\bibnamefont{Li}}, \bibnamefont{et~al.},
  \bibinfo{journal}{Ap. J.} \textbf{\bibinfo{volume}{607}},
  \bibinfo{pages}{665} (\bibinfo{year}{2004}).

\bibitem[{\citenamefont{Jackiw and Pi}(2003)}]{Jackiw}
\bibinfo{author}{\bibfnamefont{R.}~\bibnamefont{Jackiw}} \bibnamefont{and}
  \bibinfo{author}{\bibfnamefont{S.-Y.} \bibnamefont{Pi}},
  \bibinfo{journal}{Phys. Rev. D} \textbf{\bibinfo{volume}{68}},
  \bibinfo{pages}{104012} (\bibinfo{year}{2003}).

\bibitem[{\citenamefont{Mukunda}(1997)}]{Muku}
\bibinfo{author}{\bibfnamefont{N.}~\bibnamefont{Mukunda}}, in
  \emph{\bibinfo{booktitle}{Geometry, Fields and Cosmology}}, edited by
  \bibinfo{editor}{\bibfnamefont{B.~R.} \bibnamefont{Iyer}} \bibnamefont{and}
  \bibinfo{editor}{\bibfnamefont{C.~V.} \bibnamefont{Vishveshwara}}
  (\bibinfo{publisher}{Kluwer Academic Publishers},
  \bibinfo{address}{Dordrecht}, \bibinfo{year}{1997}).

\bibitem[{\citenamefont{Bergmann}(1947)}]{Berg}
\bibinfo{author}{\bibfnamefont{P.~G.} \bibnamefont{Bergmann}},
  \emph{\bibinfo{title}{Introduction to the theory of Relativity}}
  (\bibinfo{publisher}{Prentice-Hall}, \bibinfo{address}{New York},
  \bibinfo{year}{1947}).

\bibitem[{\citenamefont{Schr$\ddot{\mbox{o}}$dinger}(1985)}]{Schro}
\bibinfo{author}{\bibfnamefont{E.}~\bibnamefont{Schr$\ddot{\mbox{o}}$dinger}},
  \emph{\bibinfo{title}{Space-time structure}} (\bibinfo{publisher}{Cambridge
  University Press}, \bibinfo{address}{Cambridge}, \bibinfo{year}{1985}).

\end{thebibliography}
\end{document}